\documentstyle[12pt]{article}

\begin{document}

\baselineskip 13.6pt

\title{
Discussion: Byrne and Hall on Everett and Chalmers}
\author{ Lev Vaidman}
\date{}
\maketitle

\begin{center}
{\small \em School of Physics and Astronomy \\
Raymond and Beverly Sackler Faculty of Exact Sciences \\
Tel Aviv University, Tel-Aviv 69978, Israel. \\}
\end{center}

\vspace{.2cm}
\begin{abstract}
   Byrne and Hall (1999) criticized the argument of Chalmers (1996) in
  favor of the Everett-style interpretation. They claimed to show ``the
  deep and underappreciated flaw in {\em any} Everett-style
  interpretation''. I will argue that it is possible to interpret
  Chalmers's writing in such a way that most of the criticism by Byrne
  and Hall does not  apply. In any case their general criticism
  of the many-worlds interpretation is unfounded. The recent recognition that
  the Everett-style interpretations are good (if not the best)
  interpretations of quantum mechanics has, therefore, not been negated.
\end{abstract}

%\vfill\break

%\vskip 1.8cm
 \noindent
 {\bf 1. Introduction.~~~}  
 It is probably impossible to present an interpretation of  quantum
 mechanics in unambiguous way  without writing equations.  Chalmers's presentation of
 Everett-style interpretation also can be understood in different
 ways. Instead of equations Chalmers used some technical jargon of
 quantum theory, however, some words like ``substates'' have no clear
 meaning even for physicists. Byrne and Hall (BH) interpreted
 Chalmers's jargon in a way which leads to contradictions. In this
 note I will argue that by taking a more positive approach, one can see
 in Chalmers's writing a consistent (although not necessarily very
 persuasive) argument.
 
 In the second part of their paper BH claimed to show not only that
 Chalmers has failed to establish his Everett-inspired interpretation,
 but that ``anything resembling it should not be taken seriously''.
 Their first point is of a general character: if the spaces of states
 in two theories are identical but the dynamics is not, it is not
 obvious that the interpretation of these states in the two theories
 must be identical too. BH point out that this is the situation
 regarding the interpretation of quantum states in the orthodox and
 the Everett interpretations. I will argue that although their general
 argument is correct, its application is not.  There is enough
 similarity between the dynamics that makes the identification
 plausible. The second point of BH is that the Everett-style
 interpretation has less ``substantive content'' than the orthodox
 interpretation. This is because in the Everett (many-worlds)
 interpretation there is no counterpart of ``outcome probabilities'',
 the concept of the orthodox interpretation associated with a system
 in a superposition of eigenstates of some variable.  I will argue
 that the  definition of the probability of an
 outcome in the framework of the many-worlds interpretation which I
 recently proposed solves this difficulty and makes this BH criticism obsolete.

 The organization of this note is as follows. In Section 2 I will
 adopt the BH interpretation of Chalmers and will show (in a different
 from BH way) how it leads to a contradiction. In Section 3 I propose
 an alternative interpretation of Chalmers's writing which leads to a
 consistent argument. In Section 4 I critically analyze the general
 arguments of BH against the Everett-style interpretations. Finally,
 in Section 5 I summarize my defense of the many-worlds
 interpretation.

\vskip .8cm
 \noindent
 {\bf 2.    Byrne and Hall interpretation and a contradiction in the Chalmers argument.~~~} 
  The central thesis of Chalmers quoted by BH is the principle of
 {\it organizational preservation under superposition}:
 \begin{quotation}
\noindent
OPUS\break
   ``If a computation is implemented by a system in a maximal
   physical state $P$, it is also implemented by a system in a
   superposition of $P$ with orthogonal physical states.''(Chalmers, 350)
 \end{quotation}
 Consider a simple model: a computer which performs calculations in a
 classical way. If at time $t_0$ the computer receives a classical
 input (a particular punching of its keyboard), then it evolves in
 time is such a way that it is always in a ``classical'' state. This
 means that all the registers of the computer at all times are in some
 definite states (exited or not exited) i.e., not in a superposition
 of excited and not excited.  Suppose that $P$ corresponds to a
 computation of a square of a number 5, while $Q$ corresponds to a
 computation of a square of a number 10. Denote $|P(t)\rangle$ a
 quantum state of the computer at time $t$ performing the calculation
 of the square of 5, while $|Q(t)\rangle$ a quantum state of the
 computer at time $t$ performing the calculation of the square of 10.
 In the two computations at any time the registers must be in
 different states, therefore, $|P(t)\rangle$ is orthogonal to
 $|Q(t)\rangle$. Thus, according to OPUS the computer in a quantum
 state
 \begin{equation}
   \label{r+}
  |R_+(t)\rangle \equiv 1/{\sqrt 2}
 (|P(t)\rangle + |Q(t)\rangle), 
 \end{equation}
  also implements computation of the
 square of 5. The quantum state 
 \begin{equation}
   \label{r-}
  |R_-(t)\rangle \equiv 1/{\sqrt 2}
 (|P(t)\rangle - |Q(t)\rangle), 
 \end{equation}
 is orthogonal to $|R_+(t)\rangle$. BH read Chalmers in such a way
 that OPUS can be applied to $|R_+(t)\rangle$ and $|R_-(t)\rangle$,
 i.e., that the superposition $1/{\sqrt 2} (|R_+(t)\rangle -
 |R_-(t)\rangle)$ also implements computation of the square of 5. But,
 \begin{equation}
 {1\over{\sqrt 2}} (|R_+(t)\rangle - |R_-(t)\rangle)=
{1\over 2}[ (|P(t)\rangle + |Q(t)\rangle) - (|P(t)\rangle -
|Q(t)\rangle)] = |Q(t)\rangle .
\end{equation} 
The state  $|Q(t)\rangle$ corresponds to the computation of the square of
10. It corresponds to the punching of a different input, it has
different registers activated during the calculation, it has different
output. Clearly, it does not implement computation of the square of 5.

Applying this direct reading of Chalmers, BH reached somewhat
different contradiction which lead them to reject Chalmers's approach.

\vskip .8cm

 \noindent
{\bf 3. An alternative interpretation of Chalmers.~~~}
It is possible to read Chalmers in another way such
that the contradictions of the type described in the previous section do not arise. Let us make the following
modification of the OPUS principle:
 \begin{quotation}\noindent
   OPUS$'$\break ``If a computation is implemented by a system in a
   maximal physical state $P$ {\it which is not a superposition}, it
   is also implemented by a system in a superposition of $P$ with
   orthogonal physical states''(Chalmers, 350)
 \end{quotation}
This modified principle can be applied to $P$ and $Q$, but it cannot
be applied  to $R_+$ and $R_-$ and, therefore, one cannot reach the
contradiction described above as well as  the contradictions
described by BH.

One might see that OPUS$'$ is what Chalmers actually had in mind even
though he did not say it explicitly. Indeed, another way to see the
difference between OPUS (as read by BH) and OPUS$'$ is that in the
latter it is required that $P$ corresponds to a {\it single}
experience.\footnote{In principle, the quantum state corresponding to
  a particular experience have a nonzero overlap with quantum states
  corresponding to other experiences due to the tails of quantum waves
  which must exist because of the uncertainty principle. But these
  overlaps are so small that they can be neglected in the discussion.}
Chalmers's first {\it definition} of the OPUS principle is:
\begin{quotation}
  If the theory predicts that a system  in a maximal physical state $P$
gives rise to an associated maximal phenomenal state $E$, then the
theory predicts that a system in a superposition of $P$ with some
orthogonal physical states will also give rise to $E$. (Chalmers, 349)
\end{quotation}
The word ``associated'' hints that Chalmers meant that there is only
one experience (``phenomenal state $E$'' in Chalmers's notation) corresponding to 
physical state $P$.

In fact, BH saw a possibility of reading OPUS as OPUS$'$. The ``(Version
of) OPUS'' described in their section 5.2.3 is essentially OPUS$'$. They
rejected this because they understood that Chalmers denies the
existence of 
 {\it preferred basis}. BH are
correct in their criticism that without preferred basis there is no
way to distinguish between quantum state which is a ``superposition''
and a state which is not a ``superposition''.  Thus, the modification
of OPUS to OPUS$'$ cannot be done without assuming preferred basis. 

We can read Chalmers in such a way that we do not run into
inconsistency: Chalmers only objects to the claim that the
{\it mathematical} formalism of quantum mechanics, i.e. the Schr\"odinger
equation, leads to preferred basis. He cannot object to the existence
of preferred basis, but he  views it as arising from his theory of
consciousness. This reading of Chalmers is justified by the following
quotations:
\begin{quotation}
  Everett assumes that a superposed brain state will have a number of
  distinct subjects of experience associated with it, but he does
  nothing to justify this assumption. It is clear that this matter
  depends crucially on a theory of consciousness. A similar suggestion
  is made by Penrose (1989): ``... a theory of consciousness would be
  needed before the many-worlds view can be squared with what one
  actually observes'' (348)
\end{quotation}

\begin{quotation}
  ... last three strategies are all {\it indirect} strategies,
  attempting to explain the discreteness of experience by  explaining
  an underlying discreteness of macroscopic reality. An alternative
  strategy is to answer the question about experience {\it directly}. (349)
\end{quotation}

The main difficulty which  BH see in  putting together the {\it
  principle of organization invariance} together with OPUS follows
from the same misinterpretation of Chalmers. If there is no  preferred basis
then they have reasons to say:
\begin{quotation}
 ... perceptual experience is (more or less) {\em entirely illusory}. 
 When you seem to see a voltmeter needle pointing to `10' your
 perceptual experience is probably veridical: the needle (if, indeed,
 we can sensibly speak of such a thing) is not pointing to `10' or
 anywhere else.
\end{quotation}
However, accepting preferred basis, even if it is  defined by the concept of
experience itself, resolves the difficulty: the pointer does point to
`10' and in addition, in parallel worlds, to other values too.

Chalmers claims that his {\em independently motivated} theory of
consciousness {\it predicts} that even in the world which is in a
giant superposition there are subjects who experience a discrete
world. He bases his argument on ``the claim that consciousness arises
from implementation of an appropriate computation.'' Taking the model
of a simple computer presented above, we can follow (at least
approximately) his proof on p. 350.  Projection of the superposed
state on ``the hyperplane of $P$'' might mean projection of the
quantum state of the computer in a ``superposed'' state at the initial
time on the state corresponding to the input of calculating square of the
number 5 which leads to quantum states of the various registers at
later times corresponding to this calculation.  The parallel between
the calculation and experience yields the desired result, but
accepting this parallel is relying  on our experience. So, if
we read Chalmers as BH do, that he claims to {\it deduce} ``what the world
is like if the Schr\"odinger equation is all'' without the guide of our
experience, then they have a valid criticism.  However, Chalmers
admits that Schr\"odinger equation cannot be all:
\begin{quotation}
... the only physical principle needed in quantum mechanics is the
Schr\"odinger equation, and the measurement postulate and other basic
principles are unnecessary baggage. To be sure, we need psychophysical
principles as well, but we need those principles in any case, and it
turns out that the principles that are plausible on independent
grounds can do the requisite work here. (350-351)
\end{quotation}
I feel that these ``independent grounds'' are  connected with our
experience in a stronger way than one might imagine  reading Chalmers. But
this fact cannot lead to rejection of this approach as BH  claim.

\vskip .8cm

 \noindent
{\bf 4. Byrne and Hall against {\it any} Everett-style interpretation.~~~}
BH start their argument by pointing out  that
the orthodox quantum theory and the Everett interpretation formally
defined on the same ``family of state spaces'' and that the difference
is  only
in dynamics.
Then they say  that because of the difference in dynamics it 
does not follow that the  quantum state corresponding to a particular
experience in the orthodox theory  will correspond to the same belief
(if at any) in the framework of the Everett theory. 

This might be considered as a criticism of  Chalmers if one reads
him saying that Everett theory {\it predicts} what our experiences
should be, but usually this connection is {\it postulated} in
Everett-style theories. There is a strong motivation for this
postulate. The orthodox theory is defined only on a (tiny) part of the
space of all quantum states: macroscopic quantum systems cannot be in
a ``superposition states''. The dynamics of the allowed states between
quantum measurements is {\it identical} to the dynamics of the quantum
states in the Everett theory.
Let us discuss the example analyzed by BH at the end of p.385.  When a state
$\phi$ is a state of an observer who has the belief that the
measurement outcome was ``up'' in the orthodox theory, the dynamics
will tell that she will write  ``up'' in her lab-book. The
dynamics of the state $\phi$ in the Everett theory leads to the same
action. This justifies considering $\phi$ to be a ``belief vector'' in
the Everett theory  too.

BH proceed with their criticism claiming that  Everett's
interpretation has less of ``substantiative content'' because when a
quantum system is in a superposition of eigenstates with different
eigenvalues of some quantity {\bf M}, the orthodox interpretation
associates probabilities to the various outcomes, while the Everett
theory does not.

It is true that there is a difficulty with the concept of probability
in the framework of the Everett-style interpretation. The Everett
theory is a deterministic theory and it does not have a genuine
randomness of the collapse of the orthodox interpretation. A
deterministic theory might have the concept of {\it ignorance}
probability, but it is not easy to find somebody who is ignorant of
the result of a quantum experiment: it is senseless to ask what is the
probability that an observer will obtain a particular result, because
she will obtain {\it all} results for which there are a non-zero
probabilities according to the orthodox approach. It seems also
senseless to ask what is the probability of the observers in various
branches (these are persons with the same name and the same memories
about events which took place before the measurements, but who live in
different branches corresponding to the different outcomes) to obtain
various results, since obviously the probability to obtain the result
``${\rm \bf M} = m_i$'' in branch ``$j$'' is 1 if $i=j$ and it is 0 if
$i\neq j$. These are not the quantum probabilities we are looking for.

Nevertheless, there is solution for this difficulty (Vaidman 1998,
2000). The splitting into various branches occurs usually before the
time when the observers in these branches become aware of the outcome
of the measurement. (To ensure this we may ask the observer to keep
her eyes close during the measurement.) Thus, an observer in each
branch is ignorant about the outcome of the measurement and she can
(while any external person cannot!) define the the {\it ignorance}
probability for the outcome of the measurement.  She will do so using
standard probability postulate: the probability of an outcome is
proportional to the square of the amplitude of the corresponding
branch. Moreover, since observers in {\it all} these branches have identical
concept of ignorance probability and since they all are descendents of
the observer who performed the experiment, we can associate
probability for an outcome of a measurement for this observer in the
sense that this is the ignorance probability of her descendents in
various branches.

  The fact that I have used a
probability postulate here does not spoil the argument: I had to
show that substantive content of Everett interpretation is not less
than that of the orthodox interpretation. The latter has the
probability postulate as well. What was done here (and what was not
trivial from the beginning) is presenting a way which allows to {\it
  define} probability in the frame of the many-worlds interpretation.

The last argument of BH relies on their claim that Everett-style
interpretation lacks ``statistical algorithm''.  Since the ignorance probability defined above generates the same
statistical algorithm as the the orthodox theory, this argument does not
hold either.

\break
\vskip .8cm
 \noindent
 {\bf 5. Conclusions.~~~} The main claim of BH is ``that {\it any}
 Everett-style interpretation should be rejected''. The basis of their
 argument is the observation that neither Chalmers nor anybody else
 can answer the question: ``What the world is like if the Shr\"odinger
 equation is all?'' It is true that this question is much more
 difficult to answer in the framework of the Everett-style
 interpretation relative to interpretations which do not have
 multitude of worlds. ``The world is everything which exist'' is not a
 valid definition. Moreover, the Shr\"odinger equation itself cannot
 define the concept of a ``world''.  The world is the concept defined
 by conscious beings and it requires the analysis of the mind-body
 connection.  Chalmers's theory of consciousness provides an answer.
 One might argue how substantial his answer is, but even if there is
 no a detailed answer to this question today, one cannot  reject the
 Everett interpretation. It suffices  that 
 Everett's theory is consistent with what we see as our world. It is so
 superior to the alternatives from the physics point of view, because
 it avoids randomness and action at a distance in Nature (e.g., see
 Vaidman 2000), that it is still preferable in spite of the fact that
 it is less satisfactory from the philosophical point of view.
  Therefore, even if BH were able to point out a 
 the difficulty in obtaining the interpretation out of the ``bare theory''
 this would not be enough for rejecting the Everett interpretation.
Moreover, I have argued that the  BH have not presented persuasive
arguments  showing  the difficulty. Their first
 argument is that it is not obvious that the correspondence between
 quantum states and classical properties in the orthodox quantum
 mechanics can be transformed as it is to the Everett
 interpretation. This argument does not take into account 
 the similarity in dynamics which justifies the identification.
 Their other arguments rely on the well known difficulty in the
 interpretation of probability in the many-worlds interpretation 
 disregard a recently proposed solution of this difficulty (Vaidman,
 1998).

 In summary, BH were not able to show a flaw in Everett-style
 interpretations.  The temptation to appeal to the philosophy of mind
 in interpreting quantum mechanics, in particular, the idea that a
 theory of mind might help rescue from the difficulties with standard
 interpretation is still very attractive. Indeed, the Everett-style
 interpretation which says that physics is described in full by the
 Schr\"odinger equation is the most satisfactory  from the physics point
 of view. What is left is to complete Chalmers's work, i.e. to
 elaborate the connection between the quantum state evolving according to
 the Schr\"odinger equation and our experience.

  This research was supported in part by grant
 471/98 of the Basic Research Foundation (administered by the Israel
 Academy of Sciences and Humanities).

\vskip .8cm

\centerline{\ REFERENCES}
\vskip .3cm
%\footnotesize
\vskip .13cm \noindent 
Byrne, A. and Hall, N. (1999) ``Chalmers on Consciousness and Quantum
Mechanics'', {\it Philosophy of Science} 66: 370-390.

\vskip .13cm \noindent 
Chalmers, D. J. (1996)
{\it The Conscious Mind},  New York: Oxford University Press.

\vskip .13 cm\noindent
 Vaidman, L. (1998)      `On Schizophrenic Experiences of the Neutron or Why We
should Believe in the Many-Worlds Interpretation of Quantum Theory',
{\it International
  Studies in the Philosophy of Science} {\bf 12}, 245-261.

\vskip .13 cm\noindent Vaidman, L. (2000) ``The Many-Worlds
Interpretation of Quantum Theory'', {\it Stanford Encyclopedia of
  Philosophy} (temporarily in
http://www.tau.ac.il/${\!}\tiny\sim$vaidman/mwi/mwst1.html).

\end{document}